\documentclass[pre,superscriptaddress,tightenlines,twocolumn,showpacs]{revtex4-1}
\usepackage{latexsym,amsmath,amsfonts,tabularx,amssymb,bm,empheq}
\usepackage{subfigure}

\usepackage{color}
\usepackage[dvipsnames]{xcolor}
\definecolor{nblue}{RGB}{28,130,185}

\definecolor{cgreen}{RGB}{76,153,0}

\definecolor{myorange}{RGB}{245,156,74}
\usepackage{hyperref}
\hypersetup{
  colorlinks=true,
  citecolor=magenta,
  urlcolor=-myorange
}

\newcommand{\bea}{\begin{eqnarray}}
\newcommand{\eea}{\end{eqnarray}}

\newcommand{\del}{\partial}

\def\simge{\mathrel{%
   \rlap{\raise 0.511ex \hbox{$>$}}{\lower 0.511ex \hbox{$\sim$}}}}
\def\simle{\mathrel{
   \rlap{\raise 0.511ex \hbox{$<$}}{\lower 0.511ex \hbox{$\sim$}}}}

\def\simle{\mathrel{
   \rlap{\raise 0.511ex \hbox{$<$}}{\lower 0.511ex \hbox{$\sim$}}}}

\def\simge{\mathrel{%
    \rlap{\raise 0.511ex \hbox{$>$}}{\lower 0.511ex \hbox{$\sim$}}}}

\begin{document}

\title{Surprises in the $O(N)$ models: nonperturbative fixed points, large $N$ limit and multi-criticality}

\author{Shunsuke Yabunaka}
\affiliation{
Fukui Institute for Fundamental Chemistry, Kyoto University, Kyoto 606-8103, Japan
}

\author{ Bertrand Delamotte}
\affiliation{
Laboratoire de Physique Th\'eorique de la Mati\`ere Condens\'ee, UPMC,
CNRS UMR 7600, Sorbonne Universit\'es, 4, place Jussieu, 75252 Paris Cedex 05, France
}
%\author[1]{Shunsuke Yabunaka}
%\author[2] {Bertrand Delamotte}
%\affiliation[1]{
% Fukui Institute for Fundamental Chemistry, Kyoto University, Kyoto 606-8103, Japan}
% \affiliation[2]{Laboratoire de Physique Th\'eorique de la Mati\`ere Condens\'ee, UPMC,
%CNRS UMR 7600, Sorbonne Universit\'es, 4, place Jussieu, 75252 Paris Cedex 05, France
%}

%\renewcommand\Authands{ and }

\date{\today}

\begin{abstract}
 We find that the multicritical fixed point structure of the O($N$) models is much more complicated than widely believed. In particular, we find new nonperturbative fixed points in three dimensions ($d=3$) as well as at $N=\infty$. These fixed points come together with an intricate double-valued structure when they are considered as functions of $d$ and $N$. Many features found for the O($N$) models are shared by the O($N)\otimes$O(2) models relevant to  frustrated magnetic systems.
\end{abstract}

\maketitle

The  O($N$)-symmetric and Ising  statistical models have played an extremely important role in our understanding of second order phase transitions both because many experimental systems show this symmetry and because they have been the playground on which almost all the theoretical formalisms aiming at describing  these phase transitions have been developed and tested: Integrability \cite{2dIsing},  large-$N$ \cite{BrezinWallace,Zinn-Justin}, $4-\epsilon$ \cite{4-epsilon} and $2+\epsilon$ \cite{2-epsilon} expansions, conformal field theory \cite{conformal}, high and low temperature expansions \cite{pelissetto2000}, bootstrap program \cite{El-Showk}, all these methods  were born here. It is by now widely believed that everything is known about the criticality of the O($N$) models either exactly or with an accuracy that is limited only by our finite computational ability.

\begin{figure}[!t]
%\showfigures{
\includegraphics[scale=0.22]{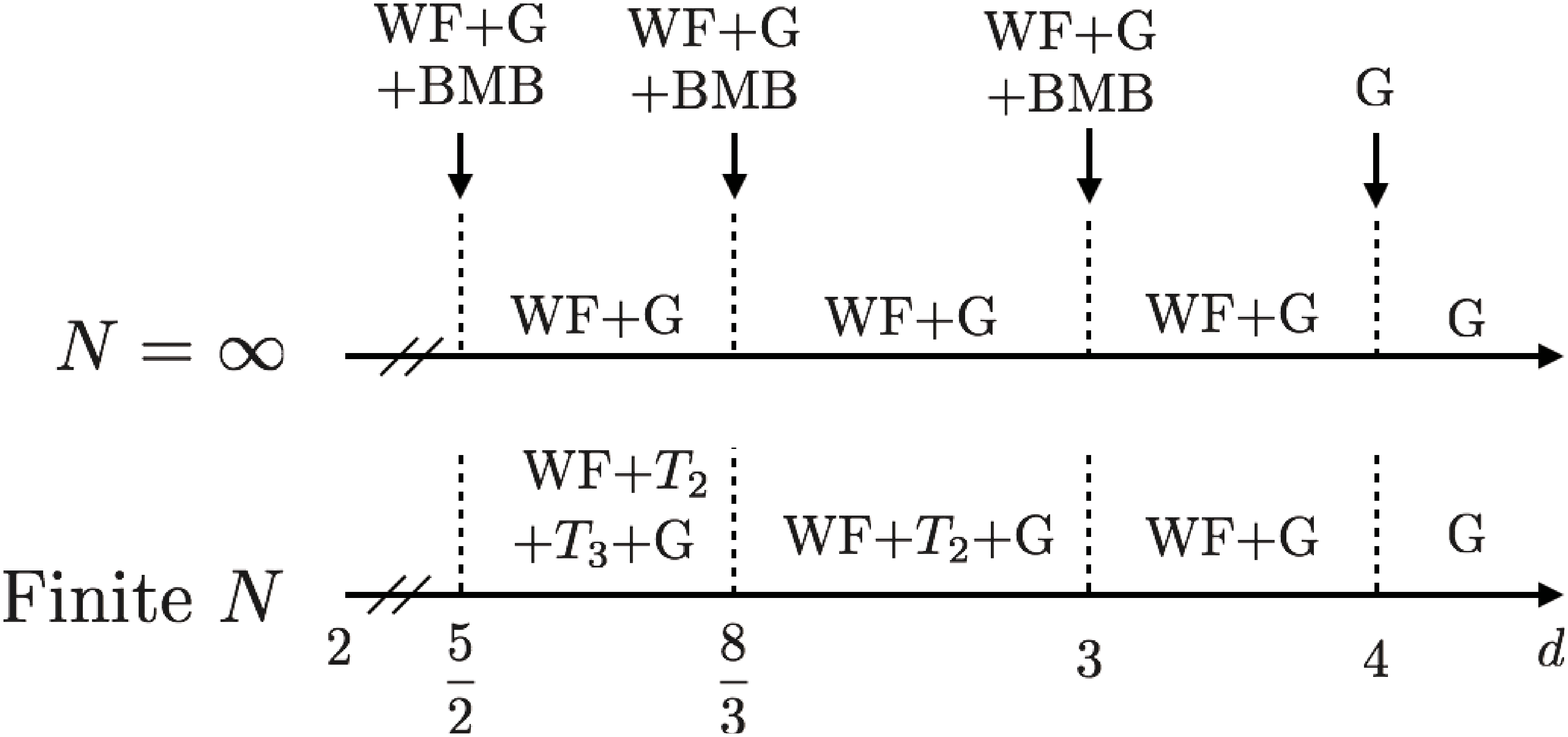}
%}
%\vspace*{0.5cm}
\caption{
Summary of the common wisdom:  Below each critical dimension $d_n=2+2/n$ a new FP emerges from the gaussian FP G.
WF stands for the Wilson-Fisher FP, $T_2$ and $T_3$ for the tricritical and tetracritical FPs. The BMB FPs exist only at $N=\infty$ and in the critical dimensions $d_{n>1}$.  
}
\label{fig:wisdom}
\end{figure}

Let us summarize the common wisdom about criticality of the O($N$) models, see Fig. \ref{fig:wisdom}, because this is what we want to challenge in this Letter \cite{Zinn-Justin}. Let us start in infinite dimension where the mean-field approximation  is exact. Lowering the dimension $d$ down to $d=4$,  the critical exponents remain those of the mean-field approximation because large scale fluctuations are Gaussian-like. This means that the only infrared fixed point (FP) of the renormalization group (RG) flow with one unstable eigendirection (1-unstable) is the Gaussian FP (G) for $d\ge4$. Since the potential part of the hamiltonian of the O$(N)$ model can only involve $({\boldsymbol{\varphi}}^2)^n$ terms, each time the dimension decreases enough for such a term to become relevant around G, that is, becomes perturbatively  renormalizable, a new nontrivial FP emerges from G. For instance, in $d=4-\epsilon$, the $({\boldsymbol{\varphi}}^2)^2$ term becomes relevant at G and a new FP, called the Wilson-Fisher FP (WF), appears. It drives the  second-order phase transition of the O($N$) models in $d<4$  and is 1-unstable while G becomes 2-unstable. The $({\boldsymbol{\varphi}}^2)^3$ term becomes relevant in $d=3-\epsilon$ and a  nontrivial 2-unstable FP emerges from  G that becomes 3-unstable. This scenario repeats in each critical dimension $d_n=2+2/n$ below which a new  $n$-unstable multicritical FP appears that we call $T_n$. The FP $T_2$ is tricritical because it lies in the coupling constant space
on the boarder separating the domain of second order and  first order phase transitions. The common wisdom is that all the $T_n$ FPs can be followed by continuity in $d$ down to $d=2$ {\sl for all values of} $N$. This is corroborated by the fact that in the Ising case ($N=1$), it has been rigorously proven that indeed all the $T_n$ exist in $d=2$ and are nontrivial \cite{Zamolodchikov}. Because of Mermin-Wagner theorem, the situation is physically different for $N\ge2$ but at least $T_2$ can be followed smoothly from $d=3-\epsilon$ down to $d=2$ for $N=2,3$ and 4 \cite{Codello}. Notice that the $N=2$, $d=2$ case is peculiar because topological defects can trigger in this case a finite-temperature phase transition. 

At $N=\infty$, exact results can be derived such as a closed and exact RG flow equation for the Gibbs effective potential \cite{Tetradis}. The common wisdom is that at $N=\infty$ and in generic dimensions $2<d<4$, the only nontrivial and nonsingular FP is WF which is simple to obtain after an appropriate rescaling by a factor $N$ \cite{dattanasio}. Its nonsingular character means that it is a regular function of the field. The limit $N=\infty$ is in fact peculiar because  in all the $d_{n}$ with $n\ge2$, and only in these dimensions, there also exists a line of FPs. In $d=3$, this line corresponds to tricritical FPs sharing all the same (trivial) critical exponents. This line starts at G and terminates at the Bardeen-Moshe-Bander (BMB) FP whose effective potential is nonanalytic  at vanishing field, see Fig. \ref{fig:wisdom} \cite{Bardeen-Moshe-Bander, David, Omid, Mati2017}.

It is surprising that this common wisdom about the O$(N)$ models raises a simple paradox that, to the best of our
knowledge,  has remained unnoticed up to now.

Let us first assume that for the O$(N)$ models, 
the exact RG flow equation of the Gibbs free energy $\Gamma$ -- also called effective action -- is  continuous  in $d$ and $N$. 
Then, assuming moreover that the FPs $\Gamma^*$ of these flows are well-defined functions of
$d$ and $N$,  they must also be continuous functions of these parameters and can therefore be followed smoothly in the $(d,N)$ plane. For constant fields, the functional $\Gamma^*[\boldsymbol{\phi}]$ reduces to the effective potential $U^*(\boldsymbol{\phi})$. If $U^*$ can be Taylor expanded: $U^*(\boldsymbol{\phi})=\sum_m g_m^* (\boldsymbol{\phi}^2)^m$ with $\boldsymbol{\phi}=\langle\boldsymbol{\varphi}\rangle$, the smoothness of $\Gamma^*$ as a function of $d$ and $N$ implies that of the $g_m^*$ which can therefore be followed continuously  along a given path of the $(d,N)$ plane.
Notice that we do not need in the following to expand $U^*$  and we indeed do not expand it. However, the same continuity argument can be used on the function $U^*$ itself rather than on its couplings.

\begin{figure}[!t]
%\showfigures{
\includegraphics[scale=0.25]{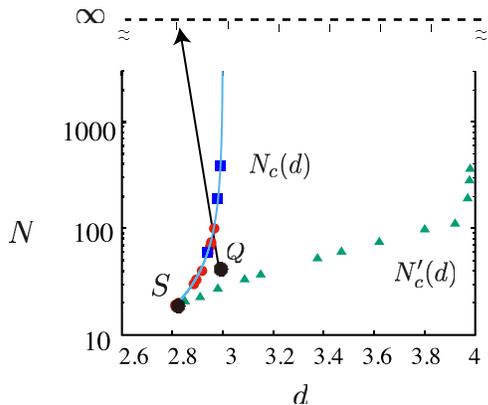}
%}
%\vspace*{0.5cm}
\caption{The two curves $N_c(d)$ and $N_c'(d)$ respectively defined by $T_2=C_3$ and $C_2=C_3$ and the  curve $3.6/(3-d)$.
$N_c(d)$ is calculated with the LPA (red circles) and at order 2 of the derivative expansion (blue squares). We show a path joining the point Q located at $(d=3^-, N=40)$ to the point at $N=\infty$ and $d=2.8$. %\textcolor{red}{I would prefer having smaller red dots, blue squares and green triangles. I think that the figure is likely to be more readable. Try and decide...}
}
\label{fig:N_c}
\end{figure}

Let us now consider for instance the tricritical FP $T_2$. The paradox appears when we try to follow smoothly $T_2$
 from a point in the $(d,N)$ plane where we know  from perturbation theory that it exists to  a point  where, according to the common wisdom, it is believed not to exist. We consider for instance the path shown in Fig. \ref{fig:N_c}
 starting at $Q$ in $d=3^-$ and $N=40$ and going to $N=\infty$ in $d=2.8$.
  How can we solve the apparent contradiction that $T_2$ should evolve continuously and that it exists at one end of the path, that is, in $Q$, and not at the other end? The simplest solution is that either $T_2$ disappears before reaching $N=\infty$ or it becomes singular at $N=\infty$. We shall see in the following that both these possibilities are indeed realized depending on the path followed to reach $N=\infty$. In particular, we shall see that there exists a line $N_c(d)$ (or equivalently $d_c(N)$), see Fig.\ref{fig:N_c}, such that when $T_2$ is followed along a path that crosses this line -- such as the path shown in Fig. \ref{fig:N_c} that starts in $Q$ --  it collapses with another FP on the line $N_c(d)$ and disappears.  This is why $T_2$ is not found at $N=\infty$ for $d<3$. And the paradox is now clear: According to the common wisdom, no known FP is available for collapsing with $T_2$. We must therefore conclude that the common wisdom yields  an incomplete picture and that there is a new FP -- that we indeed find and call $C_3$ -- with which $T_2$ collapses on $N_c(d)$. Part of the solution to the paradox above is that $C_3$ is nonperturbative: It cannot emerge from G in any upper critical dimension because 
the stability of G in the O($N)$ models is well-known for all $d$ and $N$ from perturbation theory. This is why $C_3$ has never been found previously. Some natural questions are then: What is the stability of $C_3$? Does it exist in $d=3$ for some values of $N$? 
Is it the only nonperturbative FP of the O($N$) models? Since, most probably, it does not appear alone, where does it appear and together with which other FP? Does it exist in the large-$N$ limit
and why is it not found in the usual  $1/N$ expansion \cite{BrezinWallace,Zinn-Justin,dattanasio}?
It is the aim of this Letter to provide a first study of these different questions. 

The method of choice for studying FPs beyond perturbation theory is the nonperturbative (also called functional) 
renormalization group (NPRG) which is the modern implementation of Wilson's RG. It allows us to device accurate approximate RG flows.
The NPRG is based on the idea of integrating
fluctuations step by step \cite{PhysRevB.4.3174}. In its modern version, it
is implemented on the Gibbs free energy $\Gamma$ \cite{wetterich91,wetterich93b,Ellwanger,Morris94}. A
one-parameter family of models indexed by a scale $k$ is
thus defined such that only the rapid fluctuations, with
wavenumbers $\vert q\vert > k$, are summed over in the partition
function ${\cal Z}_k$. The decoupling of the slow modes ($\vert q\vert < k$)
in ${\cal Z }_k$ is performed by adding to the original O($N$)-invariant $(\boldsymbol{\varphi}^2)^2 $ hamiltonian
$H$ a quadratic (mass-like) term which is nonvanishing
only for these modes:
\begin{equation}
 {\cal Z}_k[\boldsymbol{J}]= \int D\boldsymbol\varphi_i \exp(-H[\boldsymbol\varphi]-\Delta H_k[\boldsymbol\varphi]+ \boldsymbol{J}\cdot\boldsymbol\varphi).
\end{equation}
with
$\Delta H_k[\boldsymbol\varphi]=\frac{1}{2}\int_q R_k(q^2) \varphi_i(q)\varphi_i(-q)$
-- where, for instance, $R_k(q^2)=\alpha \bar Z_k {q^2}({\exp({q^2/k^2})-1})^{-1}$ with $\alpha$ a real parameter and $\bar Z_k$ the field renormalization --
and $\boldsymbol{J}\cdot\boldsymbol\varphi=\int_x J_i(x) \varphi_i(x)$.
 The $k$-dependent  Gibbs free energy $\Gamma_k[\boldsymbol\phi]$
is defined as  the (slightly modified) Legendre transform of $\log  {\cal Z}_k[\boldsymbol{J}]$:
\begin{equation}
\label{legendre}
 \Gamma_k[\boldsymbol\phi]+\log  {\cal Z}_k[\boldsymbol{J}]= \boldsymbol{J}\cdot\boldsymbol\phi-\frac 1 2 \int_q R_k(q^2) \phi_i(q)\phi_i(-q).
 \end{equation}
with $\int_q=\int d^dq/(2\pi)^d$.
The exact RG flow equation of $\Gamma_k$ reads \cite{wetterich93b}:
\begin{equation}
\label{flow}
\partial_t\Gamma_k[\boldsymbol\phi]=\frac 1 2 {\rm Tr} [\partial_t R_k(q^2) (\Gamma_k^{(2)}[q,-q;\boldsymbol\phi]+R_k(q))^{-1}]
\end{equation}
where $t=\log(k/\Lambda)$, ${\rm Tr}$ stands for an integral over $q$ and a trace over group indices and $\Gamma_k^{(2)}[q,-q;\boldsymbol\phi]$ is
the matrix of the Fourier transforms of the second functional derivatives of $\Gamma_k[\boldsymbol\phi]$ with respect to $\phi_i(x)$
and $\phi_j(y)$. 

\begin{figure}[!t]
%\showfigures{
\includegraphics[scale=0.22]{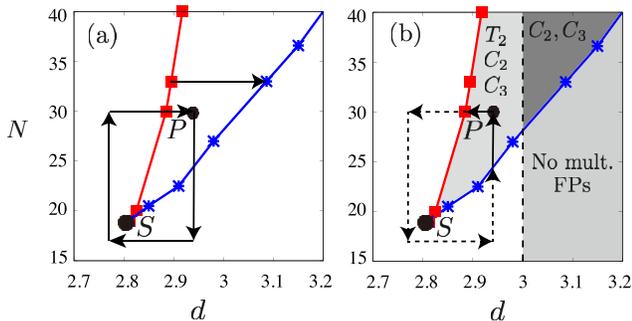}
%}
%\vspace*{0.5cm}
\caption{Singular point $S$ and the two lines $N_c(d)$ (red squares) and $N_c'(d)$ (blue stars). 
Starting from $P$, the FP $T_2$ is followed along a clockwise (left) or anti-clockwise (right) closed path surrounding $S$.
On the clockwise path, $T_2$ becomes  $C_2$ after a full rotation. On the anti-clockwise path, $T_2$ collides with $C_3$ on $N_c(d)$ and disappears. It actually becomes complex-valued and remains so all along the dashed path. On $N_c'(d)$ it becomes  real again  but is now $C_2$. 
The  path  joining $N_c(d)$ and $N_c'(d)$ at fixed $N=33$ is also shown in panel (a).In panel (b), we indicate which (real) multicritical FPs exist in each region. In the white region, there is only one multicritical FP with two directions of instability that can be continuously followed from both $T_2$ and $C_2$ depending on the path followed. 
%\textcolor{red}{PRL requires to call a the left panel and b the right one. Please correct. Those who have read the article did not understand in these figures  the symbols T $\to$ T and $C_2\to C_2$. I propose to simply eliminate them. Finally, the y axis should labelled $N$.}
}
\label{fig:closed-path}
\end{figure}

For the systems we are interested in, it is impossible to solve
Eq.~(\ref{flow}) exactly and we therefore have recourse to  approximations. The most appropriate nonperturbative  approximation 
consists in expanding $\Gamma_k[\boldsymbol\phi]$ in powers of $\nabla\boldsymbol\phi$ \cite{canet03,canet05,kloss14,delamotte04,benitez08,canet04,tissier10,tisser08,canet16,leonard15}.
At order two of the derivative expansion,
$\Gamma_k$ reads:
\begin{align}
   \begin{split}
 \Gamma_k[\boldsymbol\phi] = \int_{x} &\left(\frac{1}{2} Z_k(\rho)(\nabla \phi_i)^2 + \frac{1}{4}Y_k(\rho)(\phi_i\nabla \phi_i)^2 \right.\\
 &\left.\ \ + U_k(\rho)\vphantom{\frac{1}{4}}+O(\nabla^4)\right).      
   \end{split}
   \label{ansatz-order2}
\end{align}
where $\rho= \phi_i\phi_i/2$. Within this approximation, all critical exponents are accurately computed for all $d$ and $N$. The LPA'  (Local Potential Approximation') is a simpler approximation consisting in setting in Eq. (\ref{ansatz-order2}): $Y_k(\rho)=0$ and $Z_k(\rho)=\bar Z_k$, a field-independent field renormalization.  From $\bar Z_k$  is derived the running anomalous dimension $\eta_t=-\partial_t \log \bar Z_k$ that converges at the FP to the anomalous dimension $\eta$. The LPA consists in setting $\bar Z_k=1$ which implies $\eta=0$. The RG flow is one-loop exact in the $\epsilon=4-d$  (or $\epsilon=3-d$ for $T_2$) expansion for both the LPA and LPA' and is also one-loop exact for the LPA' for $N>1$,  in the  $\epsilon'=d-2$ expansion.  Most importantly for what follows, even within the LPA, the flow of the effective potential $U_k$ is exact at $N=\infty$. We give the flow of the effective potential $U_k$ for any $N$ at the LPA in the Supplemental Material.
%Using dimensionless and renormalized quantities, denoted by a tilde, it reads (see Supplemental Material for more details):
%\begin{align}
%   \begin{split}
% &\partial_t \tilde{U}_t= -d \tilde{U}_t +(d-2) \tilde\rho \tilde{U}_t' -2 %\int dy\, y^{d/2+1}  r'(y).  \\
%& \left(\frac{1}{y(1+r(y))+ \tilde{U}_t'+2\tilde\rho \tilde{U}_t''}+\frac{N-1}{y(1+r(y))+ \tilde{U}_t'}   \right)
% \end{split}
%  \label{flow-potential}
%\end{align}
% with $y=q^2/k^2$, $R_k(q)=q^2 r(y)$. 
%Notice that $\eta$  will always be small in the following which makes the LPA and LPA' accurate for our study.

%\begin{figure}[!t]
%\showfigures{
%\centering
%\includegraphics[scale=0.25]{fig3-629.eps}
%}
%\vspace*{0.1cm}
%\caption{$N'_c(d)$ calculated with LPA. On this curve $C_2=C_3$. }
%\label{fig:N'(d)-LPA}
%\end{figure}

We have numerically integrated the fixed point equation for the effective potential: $\partial_t \tilde U^*=0$, Eq. (S. 3), at the LPA and LPA'. As expected, we  find $T_2$ for any $N$   emerging from G in $d=3^-$. For sufficiently small values of $N$, typically $N<19$, we find that we can follow this FP down to $d=2$ using the LPA'. For   $N>19$, we find that by decreasing $d$ at fixed $N$, $T_2$ disappears in a dimension $d_c(N)$ by collapsing with a 3-unstable FP that we call $C_3$ as already explained above, see Figs. \ref{fig:N_c} and \ref{fig:closed-path}. We find that the line $N_c(d)$ is asymptotic to the $d=3$ axis, see Fig.\ref{fig:N_c}, as expected for the disappearance of $T_2$ just below $d=3$ at large $N$. A very good fit of the $N_c(d)$ curve is  $3.6/(3-d)$, see Fig. \ref{fig:N_c}.We note that this result is fully consistent with six-loop calculations performed within the $\epsilon=3-d$ expansion, see  Pisarski \cite{Pisarski} and Osborn \cite{Osborn}. Within this $\epsilon$-expansion, these authors found that at leading order in $1/N$, $T_2$ can exist only when 
$N \epsilon<36/\pi^2\simeq 3.65$ which is our bound $N_c(d)$ up to the numerical uncertainty on the prefactor 3.6 of our fit above. While this bound has been interpreted as  the radius of convergence of the $\epsilon$-expansion at large $N$ \cite{Osborn}, our results show that it is the location of the coalescence of $T_2$ with $C_3$.

We have checked that the picture above is quantitatively stable when we go from the LPA to the order two of the derivative expansion, Eq. (\ref{ansatz-order2}), see Fig. \ref{fig:N_c}. This is completely consistent with the fact that  $\eta$  is very small on the curve $N_c(d)$ for $N$ sufficiently large and decreases at large $N$ which makes the LPA flow of $U_k$ exact at $N=\infty$. For instance, for $N=40$, we find $d_c(40)=2.924$ and in this dimension, $\eta=1.7\ 10^{-3}$.  Thus, although we have no rigorous proof, we can safely claim that the existence of $C_3$ is doubtless and that the curve $N_c(d)$ approaches $N=\infty$ when $d\to3$. We show the $T_2=C_3$ FP potential shape on $N=N_c(d)$ in the Supplemental Material. It is a regular function of $\rho$ at $N=\infty$, which is not the case for the BMB FP, which shows a cusp. 

Let us now follow $C_3$ by increasing $d$. We choose for instance $N=33$ and we follow the path shown in Fig. \ref{fig:closed-path}.a starting at $d_c(N=33)= 2.90$. We find that $C_3$ exists on this path up to $d=$3.09 which shows that a nonperturbative FP can exist in $d=3$. In $d=$3.09, it collapses with
a 2-unstable FP, that we call $C_2$ and both these FPs do not exist for $d>3.09$. The  FP $C_2$ cannot be $T_2$ because  $T_2$ does not exist above $d=3$. 
By changing the value of $N$, we generate a line where $C_3=C_2$ that we call $N_c'(d)$, see Figs. \ref{fig:N_c} and \ref{fig:closed-path}.

We find two interesting features of the curve $N_c'(d)$. First, the two curves $N_c(d)$ and $N_c'(d)$ meet in a  point, that we call $S$, located at $(d=2.81, N=19)$, see Figs. \ref{fig:N_c} and \ref{fig:closed-path}. This means that right at $S$:  $T_2=C_3=C_2$. We also find that $S$ is a singular point:  If we follow smoothly $T_2$  around a closed loop containing $S$ starting for instance  at $P=(d=2.94,N=30)$, see Fig. \ref{fig:closed-path}, we do not come back at $T_2$. More precisely, starting from $P$ and following an anti-clockwise closed path as in Fig. \ref{fig:closed-path}.b, $T_2$ collides on the line $N_c(d)$ with $C_3$ and disappears. More precisely, it becomes complex. On the contrary, following the same path clockwise, $T_2$ does not collide with any FP but becomes $C_2$ after a full rotation around $S$. This is why we have claimed above that the
fate of $T_2$ when $N\to\infty$ depends on the path followed. In the Supplemental Material, we give a toy model in terms of the roots of a cubic equation that shows how $T_2$ can become $C_2$ when it is continuously followed  along a closed path surrounding $S$. From a purely mathematical  point of view, the continuity argument for following smoothly the FPs everywhere in the $(d,N)$ plane and exhibiting the double-valued structure of $T_2$ and $C_2$ makes sense only after allowing the FPs to be complex-valued (or, in a Taylor expansion, the $g_m^*$ to be complex). For instance, let us again consider Fig. \ref{fig:closed-path}.b. We start at  $P$  with $T_2$ which is very close to G. Beyond the line $N_c(d)$, 
$T_2$ becomes complex. It becomes real again when the path crosses $N_c'(d)$ and it is then $C_2$ which is far from G. If we go on following the same path, $C_2$ remains real all the way but after the second full rotation, it is  $T_2$ again.

The second interesting feature of the curve $N_c'(d)$ is that it also becomes vertical at large $N$ while being this time asymptotic to the $d=4$ axis, see Fig. \ref{fig:N_c}. We therefore conclude that most probably $C_3$ exists at $N=\infty$ everywhere for $d\in]3,4[$ and $C_2$ for $d\in]2,4[$.  However, we also find that for larger and larger  $N$ in $d>3$, the FP potentials of $C_2$ and $C_3$ become steeper and steeper at $\rho=0$ which indicates the presence of a singularity at the origin in their FP potential or its derivatives. The second derivative of the two potentials with respect to $\rho$ becomes also discontinuous at a point $\rho \ne 0$ in the large $N$ limit. These singularities are a possible explanation of the fact that these two fixed points were not found previously in large $N$ analyses\cite{Tetradis, dattanasio,Bardeen-Moshe-Bander, David, Mati2017}.  Using the LPA', we have checked that the line $N_c'(d)$ is only slightly modified compared to the LPA results because $\eta$ is small all along this line. It makes us confident that the overall picture above is not an artefact of our truncations.

\begin{figure}[!t]
\includegraphics[scale=0.25]{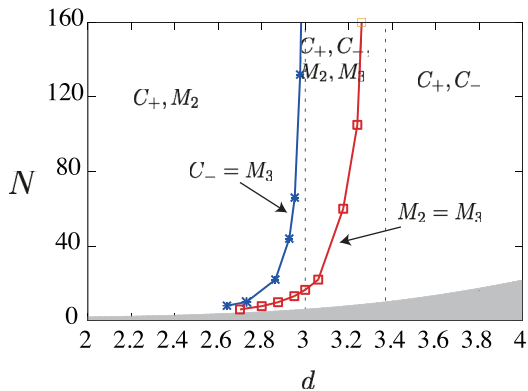}
%\vspace*{0.5cm}
\caption{
 O($N)\otimes$O(2) model. In the gray region, starting in $d=4$ at $N=21.8$, no FP at all is found. Above this region and for $d$ close to 4, both the critical $C_+$ and the tricritical $C_-$ FPs are found. The  line on the right joining the squares indicates the region where two nonperturbative FPs, $M_2$ and $M_3$, appear. On the line joining the crosses, $C_-$ and $M_3$ collapse. In each region, we indicate the FPs that are present.   
}
\label{fig:diagram2}
\end{figure}

The double-valued character of the FPs exhibited above concerns only $C_2$ and $T_2$ and we could wonder whether the same thing occurs for $C_3$. We have indeed found two other nonperturbative FPs that are 3- and 4-unstable, two analogues of the curves $N_c(d)$ and $N_c'(d)$ where these FPs show up and annihilate as well as a singular point $S'$ where the two curves meet and that shares many similarities with $S$. It is of course tempting to imagine that this kind of structure  repeats for the 4-unstable FP found that  itself involves a 5-unstable FP and so on and so forth.

%The existence of FPs at $N=\infty$ showing singular  potentials  is of course  surprising but is not completely new since the BMB FP, which is tricritical in $d=3$, shows also a cusp. The nontrivial features here are on one hand the existence of the singular FPs $C_2$ and $C_3$ in generic dimensions and, on the other hand, that these FPs also exist for generic values of $N$ and that one of them, $C_3$, is able to  annihilate with $T_2$.

A natural question is whether the intricate FP structure presented above is specific to the O($N$) models or is generic. To shed some light on this question, we have therefore considered the O$(N)\otimes$O(2) model which is relevant for frustrated antiferromagnetic systems \cite{Delamotte-review,Kawamura-review,Caffarel}.
The order parameter of this model is the $N\times2$ matrix $\Phi=\left(\mathbf{\boldsymbol{\varphi}}_{1},\mathbf{\boldsymbol{\varphi}}_{2}\right)$ \cite{Yosefin} and the  Hamiltonian is the sum of the usual kinetic terms and of the potential
%\begin{equation}
%H=\int d^{d}\mathbf{x}\left(\frac{1}{2}\left[\left(\partial\boldsymbol{\varphi}_{1}\right)^{2}+\left(\partial\boldsymbol{\varphi}_{2}\right)^{2}\right]+U\left(\mathbf{\boldsymbol{\varphi}}_{1},\mathbf{\boldsymbol{\varphi}}_{2}\right)\right)\label{eq:effective hamiltonian} 
%\end{equation}
 $U= r (\mathbf{\boldsymbol{\varphi}}_{1}^2+\mathbf{\boldsymbol{\varphi}}_{2}^2) + u(\mathbf{\boldsymbol{\varphi}}_{1}^2+\mathbf{\boldsymbol{\varphi}}_{2}^2)^2 +v (\mathbf{\boldsymbol{\varphi}}_{1}^2\mathbf{\boldsymbol{\varphi}}_{2}^2- (\mathbf{\boldsymbol{\varphi}}_{1}\cdot\mathbf{\boldsymbol{\varphi}}_{2})^2)$. By a suitable choice of $r$, $u$ and $v$ the symmetry is spontaneously broken down to O$(N-2)\otimes$O(2). For $N$ typically larger than 21.8, two FPs are found in $d=4-\epsilon$, a critical one, $C_+$,  that can be followed smoothly down to $d=2$ and another one, $C_-$, which is tricritical \cite{Jones, Bailin}. These FPs are also found in the large $N$ expansion in all dimensions between 2 and 4 \cite{Bailin, Gracey, Pelissetto}. However, using the LPA', we find for $C_-$ a picture which is very much similar to the O($N$) case, see Fig. \ref{fig:diagram2}: (i) There exists a line where $C_-$ collapses with a 3-unstable FP, that we call $M_3$; (ii)  this line is asymptotic to the $d=3$ axis, and (iii) $M_3$ appears on another line together with a 2-unstable FP that we call $M_2$ \cite{Frustrated-multicritical}.  
 %A major difference with the O($N$) case is that the curve defined by $M_2=M_3$ is not asymptotic to the $d=4$ axis but to the $d=3.36$ axis. Another difference is that the FP potentials of both $M_2$ and $M_3$ remain regular in the large $N$ limit which makes intriguing that they were not found in the $1/N$ expansion. The explanation of this fact is probably to be searched in the unusual scaling in $N$ of the couplings of these potentials. This study is under way. It is also remarkable that the two curves approach at small $N$ in much the same way as the curves $N_c(d)$ and $N_c'(d)$ do in the O$(N)$ case. Because of numerical difficulties that are very severe in the O$(N)\otimes$O(2) case, we have not been able to show  that they also meet at a singular point analogous to $S$, although this seems probable. 

To conclude we have found that the multicritical FP structure of both the O($N$) and O($N)\otimes$O(2) models is much more complicated than usually believed. In particular, we have shown that several nonpertubative FPs exist in $d=3$ that were not previously found. Although they also exist at $N=\infty$ on a finite interval of dimensions they were not found by previous direct studies of this case and this is clearly a subject that must be further studied, see however \cite{Pisarski}. The existence and role of possible singularities of the FP potential of 
$C_2$ and $C_3$  should be studied in the future as well.
It would also be interesting to study  the $d=3$ case and figure out what the basins of attraction of both $C_2$ and $C_3$ are to know whether the multicriticality of some lattice models could be described by these FPs. The NPRG, here again, is a method of choice for this study but the conformal bootstrap program could probably definitively prove/disprove the existence of the $C_2$ and $C_3$ FPs in $d=3$. We can also expect that there are other nonperturbative FPs that  collide with $T_n$ $(n=3,4,\cdots)$ as $C_2$ does with $T_2$. They are also left for future study. Finally, an intriguing question is: Could it be that 
what we have found above has for a known physical system an   impact on its criticality or multicriticality?

We acknowledge  H. Chat\'e, N. Defenu, C. Duclut, N. Dupuis, J.-M. Maillard, H. Osborn, R. Pisarski, G. Tarjus and M. Tissier for discussions and/or  advices about the manuscript.

%\textcolor{red}{Finally, we note that appearance of nonperturbative multicritical FPs between $2<d<4$ is not a special feature of the $O(N)$ models. We will report that nonperturbative multicritical FPs appear  between $2<d<4$ also in $O\left(N\right)\otimes O\left(2\right)$ models, whose order parameter is given by the $N\times2$ matrix $\left(\mathbf{\boldsymbol{\phi}}_{1},\mathbf{\boldsymbol{\phi}}_{2}\right)$ \cite{Frustrated-multicritical}. The Ginzburg-Landau-Wilson Hamiltonian reads 
%\begin{equation}
%H=\int d^{d}\mathbf{x}\left(\frac{1}{2}\left[\left(\partial\boldsymbol{\phi}_{1}\right)^{2}+\left(\partial\boldsymbol{\phi}_{2}\right)^{2}\right]+U\left(\mathbf{\boldsymbol{\phi}}_{1},\mathbf{\boldsymbol{\phi}}_{2}\right)\right).\label{eq:effective hamiltonian} \end{equation}
%Here we have defined the local potential $U\left(\mathbf{\boldsymbol{\phi}}_{1},\mathbf{\boldsymbol{\phi}}_{2}\right)$whose minima are given by $\boldsymbol{\phi}_{i}\cdot\boldsymbol{\phi}_{j}=const\times\delta_{ij}$. As shown in Fig. \ref{fig:diagram2}, apart  from the well known perturbative FPs $C_+$ (chiral FP) and $C_-$ (antichiral FP), nonperturbative multicritical FPs $M_2$ and $M_3$ appear between $2<d<3.36$ and $3<d<3.36$ at $N=\infty$, respectively. They have not yet been found to the best of our knowledge and have 2 and 3 relevant directions, respectively. We have found that, at $N=\infty$, the FP potentials are smooth functions of $\mathbf{\boldsymbol{\phi}}_{1}$ and $\mathbf{\boldsymbol{\phi}}_{2}$ without a cusp, which is not the case for $O(N)$ models. }  

\pagebreak
\widetext
% \newpage
% \onecolumngrid 

\begin{center}
\textbf{\large Supplemental Materials\\
Surprises in the $O(N)$ models: nonperturbative fixed points, large $N$ limit and multi-criticality} 
\end{center}
\section{the flow equation of the effective potential within LPA approximation}

\begin{figure}[!ht]
\centering
\includegraphics[scale=0.45]{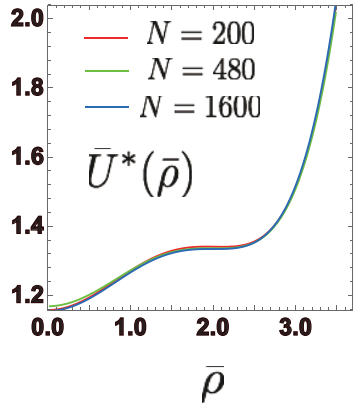}
\vspace*{0.1cm}
\caption{
The rescaled FP potential at $d=d_c(N)$ with $N=200,480$ and $1600$. We can see that the FP potential almost converges to a limiting function, which is different from the GFP.
}
\label{fig:rescaled-pot-T=C3}
\end{figure} 

We define the dimensionless  field $\tilde{\rho}$ and potential $\tilde{U}_{k}$ as
\begin{equation}
\begin{array}{l}
\tilde{\rho}=v_d^{-1} k^{2-d}\rho\\
\tilde{U}_{k}(\tilde{\rho})=v_d^{-1} k^{-d}U_{k}\left(\rho\right)
\end{array}
\end{equation} with 
\begin{equation}
v_{d}=\frac{1}{2^{d-1} d \pi^{d/2}\Gamma\left(\frac{d}{2}\right)}.
\end{equation}
Using $t=\log(k/\Lambda)$ and the dimensionless and renormalized quantities, denoted by a tilde, the flow of the effective potential $\tilde{U}_{k}$ reads at the LPA:
\begin{align}
   \begin{split}
 &\partial_t \tilde{U}_t= -d \tilde{U}_t +(d-2) \tilde\rho \tilde{U}_t' -\frac{d}{2} \int_{0}^{\infty} dy y^{d/2+1}  r'(y)  \left(\frac{1}{y(1+r(y))+ \tilde{U}_t'+2\tilde\rho \tilde{U}_t''}+\frac{N-1}{y(1+r(y))+ \tilde{U}_t'}   \right)
 \end{split}
  \label{flow-potential}
\end{align}
 with $y=q^2/k^2$, $R_k(q)=q^2 r(y)$, $\tilde{U}_t'=\del_\rho \tilde{U}_t$, $\tilde{U}_t''=\del^2_\rho \tilde{U}_t$. 
 We employed the cutoff $R_k(q)=(k^2-q^2) \theta(k^2-q^2)$ [44].%\cite{Litim}.
 This leads to $r(y)=(1/y-1) \theta(1-y)$, which is convenient for analytial treatments for LPA calculations. With this cutoff, the flow equation becomes
 \begin{align}
   \begin{split}
 &\partial_t \tilde{U}_t= -d \tilde{U}_t +(d-2) \tilde\rho \tilde{U}_t' +   \frac{1}{1+ \tilde{U}_t'+2\tilde\rho \tilde{U}_t''}+\frac{N-1}{1+ \tilde{U}_t'}. 
 \end{split}
  \label{flow-potential}
   \end{align}

\section{the $T_2=C_3$ FP potential shape on $N=N_c(d)$}
It is also interesting to notice that by rescaling the potential and the field: $U\to \bar{U} \equiv U/N$ and $\rho\to \bar{\rho} \equiv \rho/N$, the explicit factor $N$ in the LPA flow of the potential, Eq.(\ref{flow-potential}), disappears in the large $N$ limit, if we assume no singularities of $U$, $U'$ and $U''$. This implies that for large enough values of $N$, the  shape of the rescaled FP potential  is almost independent of $N$. Using this rescaling, we  find that this limit shape of $\bar {U}^*$ on the line $N_c(d)$ when $d\to3$ (or equivalently when $N\to\infty$) is clearly regular and not gaussian even though $T_2$ is closer and closer to the GFP at large $N$. This limit FP is therefore also different from the BMB FP, which shows a cusp, as shown in Fig. S1.

 This means that, for a fixed and large value of $N$,  the shape of the rescaled potential $\bar {U}^*$  changes very rapidly between $d=3^-$, where $T_2$ coincides with the GFP, and $d=d_c(N)$ where it collapses with $C_3$.

\section{A toy model of the double valued structure of $T_2$ and $C_2$}
\begin{figure}[!t]
\centering
\includegraphics[scale=0.25]{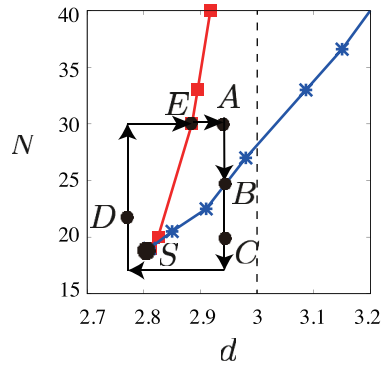}
\vspace*{0.1cm}
\caption{
A closed clockwise path around the singular point $S$.}
\label{fig:closed-path-S}
\end{figure}

\begin{figure}[!t]
\centering
\includegraphics[scale=0.3]{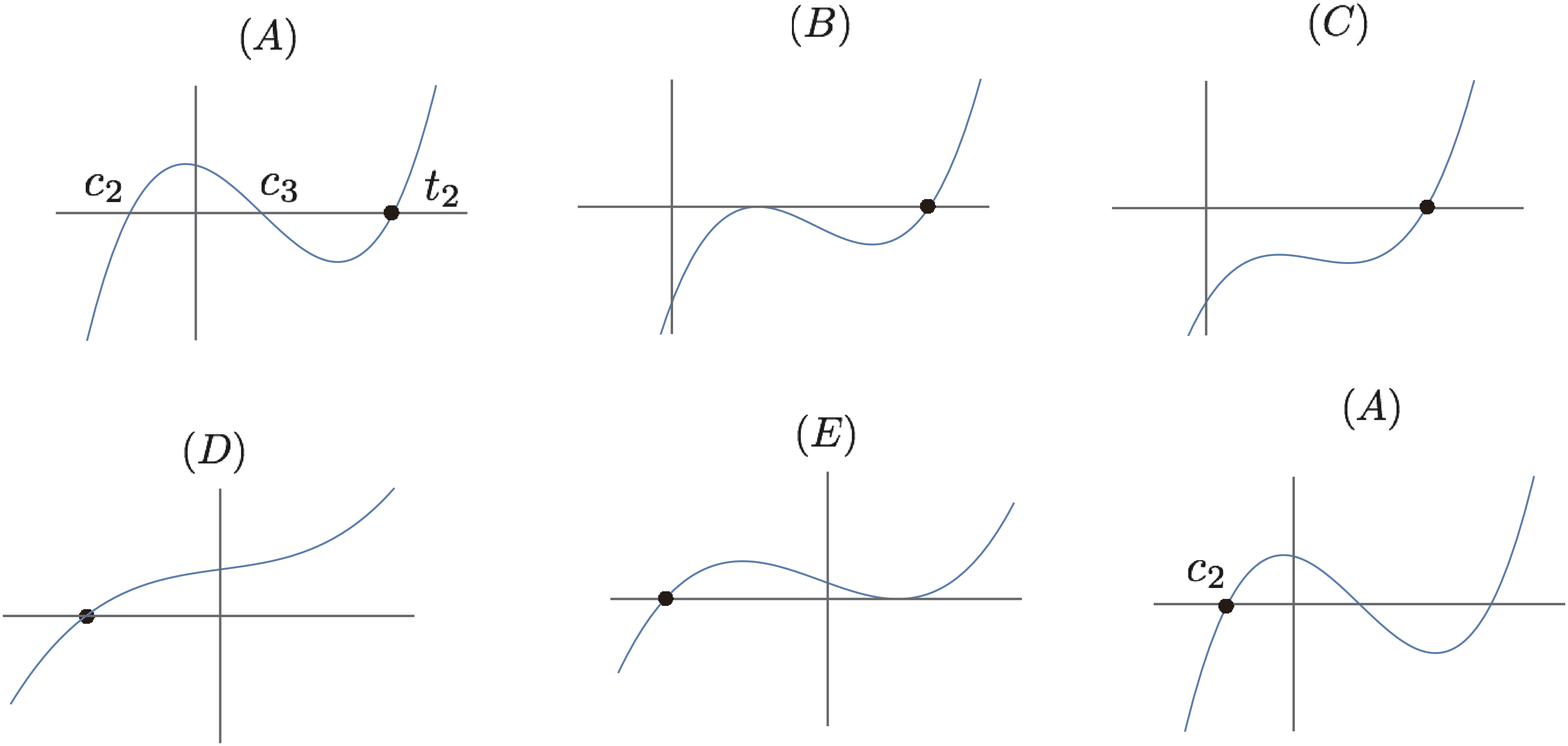}
\vspace*{0.1cm}
\caption{
Behavior of the cubic function $f(x,\theta)$ as well as its roots when $\theta$ is varied between 0 and $2\pi$. Starting from $t_2$ at (A), we follow this root by continuity all along the path, as indicated with black dots. At  $\theta=2\pi$, $t_2$ has become $c_2$. This mimics the behavior of the FP $T_2$ along the path ABCDEA in Fig. \ref{fig:closed-path-S}.}
\label{fig:toymodel-S}
\end{figure}

In this section, we give a toy model in terms of the roots of a cubic equation: $f(x,\theta)=0$, that show a similar double-valued structure as the tricritical FPs $T_2$ and $C_2$. These three roots, real or complex, depend on a parameter $\theta\in [0,2\pi]$ in a cyclic way because we assume that $f(x,0)=f(x,2\pi)$.
By analogy with our initial problem, we call $t_2, c_3$ and $c_2$ the three roots. When $\theta=0$ the three roots are assumed real and $t_2$ is the rightmost root as shown in Fig. \ref{fig:toymodel-S} (A). When 
$\theta$ is increased the two roots $c_3$ and $c_2$ are assumed to approach each other and eventually coincide as in \ref{fig:toymodel-S} (B). This corresponds to what happens in Fig.  \ref{fig:closed-path-S} where B is on the line $N_c'(d)$, that is, where the FPs $c_3$ and $c_2$ coincide. When $\theta$ is further increased, the roots $c_3$ and $c_2$ are assumed to become complex as in Figs. \ref{fig:toymodel-S} (C) and (D). In our toy model, the inflexion point which is in the lower half-plane in Fig. \ref{fig:toymodel-S} (C) is in the upper half-plane in Fig. \ref{fig:toymodel-S} (D). In Fig. \ref{fig:toymodel-S} (E) two roots become  real again as in Fig.  \ref{fig:closed-path-S} where 
$t_2=c_3$. When $\theta=2\pi$, we are back at  point A. The root $t_2$ that has been followed by continuity has therefore become $c_2$ after a full rotation which exactly corresponds to the situation indicated in Fig. 3 (a) of the main text. Notice that by further increasing $\theta$ from $2\pi$ to $4\pi$, that is, by making a second full rotation in Fig.  \ref{fig:closed-path-S}, $c_2$ would become $t_2$. 

\end{document}